# Modified Holder Exponents Approach to Prediction of the USA Stock Market Critical Points and Crashes

Yu.A Kuperin[*], R.R. Schastlivtsev [+]
Saint Petersburg State University
[*] Department of Physics, Saint Petersburg State University, #1 Ulyanovskaya Str., Saint Petersburg, 198054, Russia
E-mail: yuri.kuperin@gmail.com

[+] Department of Mathematics and Mechanics, Saint Petersburg State University, #28 University Av., Saint Petersburg, 198054, Russia
E-mail: srr2001@mail.ru

**ABSTRACT**

The paper is devoted to elaboration of a novel specific indicator based on the modified Holder exponents. This indicator has been used for forecasting critical points of financial time series and crashes of the USA stock market. The proposed approach is based on the hypothesis, which claims that before market critical points occur the dynamics of financial time series radically changes, namely time series become smoother. The approach has been tested on the stylized data and real USA stock market data. It has been shown that it is possible to forecast such critical points of financial time series as large upward and downward movements and trend changes. On this basis a new trading strategy has been elaborated and tested.

**PACS** 89.65.Gh

## 1. INTRODUCTION

Extraordinary or extreme events, which are typical for many natural and social systems belong to the so–called complex systems processes. One of the examples of such complex systems is financial market and crashes on the market are treated as extreme events. The crash on financial market is defined as noticeable decline of indexes or even separate stocks quotations in short period of time. The noticeable decline is interpreted as the price change expressed in $K\%$, where the threshold value $K$ is specific for various financial instruments and even for different time intervals. For example for DJIA (Dow Jones Industrial Average) index, which is considered in section 4.2 the value of $K$ can be accepted on the level 9% in three days. As it well known the extreme events in financial markets are very interesting for both academic researchers and investors. According to the till now prevailing paradigm the market is highly efficient, i.e. obeys the Efficiency Market Hypothesis [1]. It means that only exogenous information can be the cause of the market critical phenomena or crashes. In reality, however even very accurate study of the underlying conditions of market crashes gives no well-defined conclusion concerning what definitely provokes the crash. In that way the Efficiency Market Hypothesis and consequently any linear theory exhibits its inefficiency in market crashes analysis.

Today in the field of econophysics there are a lot of quantitative approaches mostly based on the nonlinear market behavior concept aimed to analyze and model the financial instruments dynamics. Among these approaches fractal and multifracial ones [2-6] plays the essential role. As it has been mentioned in [7-11], these approaches are very often used in market crashes predictions. Among the known multifracial techniques the special role belongs to approaches based on the study of local regularity of financial time series. Usually in order to study the local regularity various special indicators are constructed. For instance, in [12] to forecast crisis on the Brazilian stock market the Hurst exponents for Ibovespa index of the Sao Paulo Stock Exchange have been constructed. In [13] for the 1987 crash of the USA stock market prediction the pointwise Holder exponents have been used. In the paper [14] authors used the local Holder exponents for the critical events analysis of the USA stock market and currency market. One should note the research group "Groupe Fractales" (www.inria.fr), which for the first time established the idea to use the Holder exponents for detection of market crashes.

The main goal of the present paper is to elaborate on the basis of multifracial technique a novel indicator for predicting critical points of financial time series. Here the critical points are treated as trends change points, crashes or just noticeable upward and downward price changes. For the prediction of such critical points there has been elaborated a numerical algorithm for calculating the local regularity of time series. This algorithm is based on the elaboration of the novel indicator – modified Holder exponent. Forecasting by means of this indicator is based on the conjecture that before the critical event the intrinsic dynamics of financial time series radically changes. Namely, the time series becomes in some sense more regular. Qualitatively one can explain this phenomenon as universal reduction of investments horizons before approaching a critical point. In other words, traders with long investment horizons try to reduce them as much as possible in order to have time to close their position before the market crash will take place. Ultimately, it leads to the shrinking of the horizons spectrum and as a consequence to increasing the time series regularity. Thus calculating the regularity of the given financial time series at the present moment one can make an assumption about possible critical points in the future.

**2. MULTIFRACTAL ANALYSIS**

Multifractal analysis [15] is relatively recent mathematical field of study. Due to its universality, this approach is effectively used not only in the field of mathematics or physics. There are many branches of science such as linguistics, medicine, informatics etc, where the multifractal analysis is also very successful. In particular, it can be used in data mining of financial data. The numerous studies (see, e.g. [2,3,6]) have shown that financial time series possess the mulifractal properties and hence the modified Holder exponents can be calculated. In the present paper the temporal dynamics of the modified Holder exponents is the basis of proposed forecasting technique of the large market movements.

**2.1. Fractal dimension**

In the definition of fractals, the main idea by Mandelbrot [16] has been related with the notion of selfsemilarity. In other words fractals are scale invariant objects. This scale invariance may be exact or statistical. It can be shown [16] that for fractals objects the following relation is valid:

$$N(\varepsilon) \sim \varepsilon^{-D_F},$$

where $N(\varepsilon)$ is the number of cubes of the size $\varepsilon$ which form the covering of the set in question. The quantity $D_F$ is called the fractal dimension of the set. The power law dependence above leads to the scale invariance. The fractal set which is characterized by the unique universal parameter $D_F$ is called as monofractal [15].

One can define multifractal set as not uniform fractal object. It is obviously that for the whole description of multifractals it is insufficient to use the box-counting dimension $D_F$. It s necessary to introduce [15] the infinite number of the so-called generalized fractal dimensions (see, Section 2.2). Multifractal mathematical formalism exists in several versions [15-18]. Two approaches are briefly described below.

**2.2. Generalized fractal dimensions**

As in the case of fractal dimension definition let us use the box-counting method [15]. Suppose that the set in question is covered by cubes of size $\varepsilon$ and to each cube one can assign the quantity $\mu_i(\varepsilon)$ often called "mass". For example, if the set consists of isolated points the mass $\mu_i(\varepsilon)$ is the ratio of number of points in the cube with the number $i$ to the whole number of points in the set. So masses can be treated as probabilities that the point belongs to the given cube and describe relative occupancy of cubes from the covering. In case of homogeneous fractal, all cubes from the covering will contain the same number of points and hence all cubes will have the same relative occupancy. For continuum sets and multifractal measures the "masses" $\mu_i(\varepsilon)$ can be naturally introduced in each concrete case. The normalization condition $\sum_{i=1}^{N(\varepsilon)} \mu_i(\varepsilon)$ is always valid. Here $N(\varepsilon)$ is the number cubes from the covering of the set in question. The so-called partition function or statistical sum which is characterized by the real parameter $q$ is introduced as follows:

$$Z(q,\varepsilon) = \sum_{i=1}^{N(\varepsilon)} \mu_i(\varepsilon)^q, q \in (-\infty, +\infty).$$

The scaling function is defined by the relation:

$$\tau(q) = \lim_{\varepsilon \to 0} \frac{\ln(Z(q,\varepsilon))}{\ln \varepsilon}.$$

In terms of these quantities, the generalized fractal dimensions $D_q$ are introduced by the equation [19]:

$$D_q = \frac{\tau(q)}{q-1} = \lim_{\varepsilon \to 0} \frac{1}{q-1} \cdot \frac{\ln\left(\sum_{i=1}^{N(\varepsilon)} \mu_i(\varepsilon)\right)}{\ln \varepsilon}.$$

It can be shown [18], that if for the given set the generalized fractal dimensions $D_q$ do not depend on $q$, then the set is monofractal. The scaling function $\tau(q)$ for monofractal is obviously linear.

### 2.3. Multifractal spectrum: Legendre transforms

Sometimes it is more convenient to use variables which are different from $D_q$ and $q$, but are closely related with the latter. Namely let us introduce the spectrum of multifractal dimensions $f(\alpha)$ where the variable $\alpha$ is the Holder-Lipschitz exponent for multifractal set in question. The change of variables from $\tau(q)$ and $q$ to $f(\alpha)$ and $\alpha$ is given by the Legendre transforms:

$$\alpha = \frac{d\tau}{dq}, f(\alpha) = q\frac{d\tau}{dq} - \tau(q),$$

$$q = \frac{df}{d\alpha}, \tau(q) = \alpha\frac{df}{d\alpha} - f(\alpha).$$

For multifractal objects, the condition $\frac{d^2\tau}{dq^2} \neq 0$ providing the existence of the Legendre transform is valid. One can show [19], that the quantity $f(\alpha)$ is equal in fact to the Hausdorff dimension of some homogeneous fractal subset which gives the principal contribution into the statistical sum at the given $q$. Since the fractal dimension of subset is always equal or less then the fractal dimension of the whole set it leads to inequality $f(\alpha) \leq D_0$. One can also show [15], that the function $f(\alpha)$ is convex for any multifractal set. The spectrum of multifractal dimensions $f(\alpha)$ is called also as the Legendre spectrum.

There is another approach [17,18] to the spectrum of multifractal dimensions definition. Namely, consider a measure, which is distributed on the interval in arbitrary manner. Let us divide the support of the measure into equal boxes $C_\delta$ of the size $\delta$. Let $C = \{C_\delta\}$ be the set of all such boxes. Let us assume that inside of each box the measure satisfies to the relation $\mu\{C_\delta\} \approx \delta^{\alpha(C_\delta)}$, where the quantities

$$\alpha(C_\delta) = \frac{\log \mu(C_\delta)}{\log \delta}$$

are the so-called coarse-grained Holder exponents of the measure singularities. Let $N_\delta(\alpha,\varepsilon) = \#\{C_\delta : \alpha(C_\delta) \in (\alpha-\varepsilon, \alpha+\varepsilon)\}$, here (#) is the number of non empty boxes which measure is characterized by exponents from the interval $(\alpha-\varepsilon, \alpha+\varepsilon)$. Then the coarse-grained multifractal spectrum of large deviation [18,19] is given by equation

$$f_g(\alpha) = \lim_{\varepsilon \to 0} \limsup_{\delta \to 0} \frac{\log N_\delta(\alpha,\varepsilon)}{\log 1/\delta}.$$

The last expression coincides formally with the definition of the capacity of a set if this set consists from boxes with the fixed value of the measure singularity exponent. The term coarse-grained is related with the finite precision of estimation of $\alpha$. The definition of $f_g(\alpha)$ is connected with the large deviation theorem which gives the probabilistic interpretation for the multifractal spectrum. Namely, the probability to find that $\alpha(C_\delta) \approx \alpha$ behaviors at small $\delta$ as follows:

$$N_\delta(\alpha,\varepsilon)/N_\delta = P_\delta[\alpha(C_\delta) \approx \alpha] \approx \delta^{d_\tau - f_g(\alpha)}.$$

Here $N_\delta$ is the total number of boxes covering the support of the measure and $d_\tau$ is the topological dimension of the measure support. Thus, the pair $\{\alpha, f_g(\alpha)\}$ describes the structure of an arbitrary measure. If the measure is multifractal, then its spectrum $f_g(\alpha)$ is a convex curve.

## 3. THE HOLDER EXPONENTS DEFINITION AND CALCULATIONS

As it was mentioned above, there are several approaches to the Holder exponents definition. One of these approaches is based on generalized fractal dimensions definition. Then by means of Legendre transform, one can define the local Holder exponents [15]. Another approach uses the notion of fractal measure and defines on this basis the coarse-grained Holder exponents of the measure singularities [17]. In the present paper, the approach to the definition of the Holder exponents is based on the introduction of so-called Holder semi-norm. In more details, this approach is described below.

Definition 1. *Function $f(x)$, defined on the domain $E \subset R$ satisfies on the set $E$ the Holder condition with the exponent $\alpha$, where $0 < \alpha \leq 1$, and with the coefficient $A$, if*

$$|f(x) - f(y)| \leq A|x-y|^\alpha$$

*for any $x, y \in E$.*

The quantity

$$C_a = \sup_{x,y \in E} \frac{|f(x) - f(y)|}{|x-y|^\alpha}, \quad 0 \leq \alpha \leq 1, \tag{1}$$

is called the Holder $\alpha$-semi-norm of the bounded function defined on the set $E$.

Let $f(x)$ satisfies the Holder condition with the exponent $\alpha$ on the set $E$. Then one can easily show that

$$\begin{cases} C_\beta = 0, & \text{if } \beta < \alpha, \\ 0 < C_\beta < +\infty, & \text{if } \beta = \alpha, \\ C_\beta = +\infty, & \text{if } \beta > \alpha. \end{cases}$$

The behavior of $C_\beta$ depending on $\beta$ is shown in Fig. 1 by the dotted line. From the definition of $C_\beta$ it follows that there is a unique value of $\beta$, such that

$$0 < C_\beta < +\infty,  \qquad (2)$$

and such $\beta$ coincides with the Holder exponent $\alpha$. This property gives the algorithm of the Holder exponent $\alpha$ calculations. In accordance with equation (1) one calculates $C_\beta$ for all $0 \le \beta \le 1$. Further the value of the coefficient $\beta$, which satisfies the condition (2) is chosen. This value of $\beta$ is the required Holder exponent $\alpha$. It is obvious that numerical realization of the algorithm described above is embarrassed by difficulties connected with the control of the right hand side of condition (2). The numerical realization of the algorithm for time series is described below.

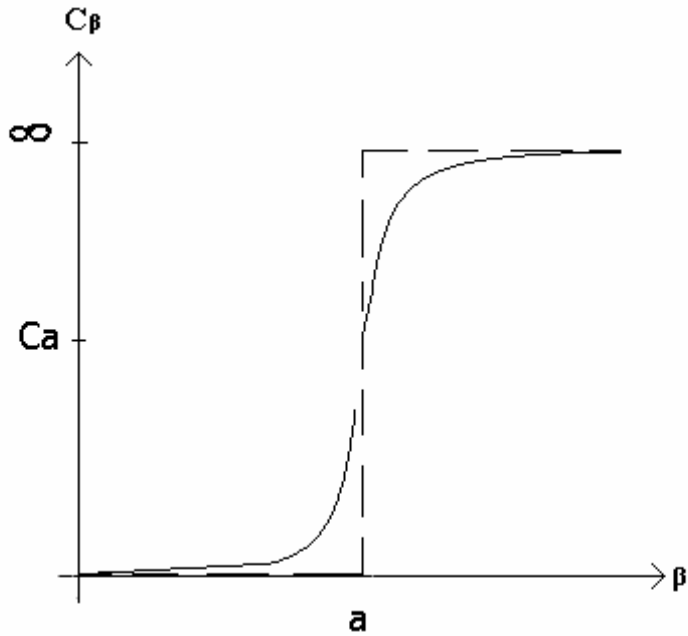

Fig. 1. The behavior of $C_\beta$ versus $\beta$ in continuous (dotted line) and discrete (solid line) cases.

Let variable $t$ takes the discrete values $t_i$ from the interval [0,1]: $t_i = i/N$, $i = \overline{1..N}$. Let $X_i$ be the value of function $f$ at points $t_i$. In that way one obtains the discrete time series $\{X_i\}_{i=1}^N$. For the algorithm realization, it is also necessary to define the increment of the parameter $\beta$. Toward this end let us fix the natural number $n$, define the increment as $\Delta\beta = 1/n$ and introduce the quantities $\beta_k = k/n$, where $k = \overline{0..n}$. In the discrete case, the equation (1) takes the form:

$$C(\beta) = \max_{i,j=1..N, i \ne j} \frac{|X_i - X_j|}{|t_i - t_j|^\beta} , \quad \beta = \beta_k, k = \overline{1..n}. \qquad (3)$$

One has to study the behavior of $C(\beta_k)$ for different $k$. For that let us introduce the value

$$\overline{k}: |\alpha - \beta_{\overline{k}}| = \min_{k=1..n} |\alpha - \beta_k|,$$

where $\alpha$ is the Holder exponent of the function $f$. It means that $\beta_{\overline{k}}$ is the best approximation of $\alpha$ among all $\beta_k$. Expressing $C(\beta_k)$ in terms of $C(\beta_{\overline{k}})$ one obtains

$$C(\beta_k) = C(\beta_{\overline{k}}) \max_{i,j=1..N, i \neq j} \frac{1}{|t_i - t_j|^{\beta_k - \beta_{\overline{k}}}}.$$

If $k < \overline{k}$, that is $\beta_k < \beta_{\overline{k}}$, then for sufficiently large $n$ and $N$ the value of

$$\max_{i,j=1..N, i \neq j} \frac{1}{|t_i - t_j|^{\beta_k - \beta_{\overline{k}}}},$$

and hence the value of $C(\beta_k)$ will be close to zero. The smaller the value of $k$ one takes, the closer to zero the quantity $C(\beta_k)$ becomes. In a similar manner at $k > \overline{k}$ the value of $C(\beta_k)$ will be much greater then 1 for sufficiently large $n$ and $N$. The quantity $C(\beta_k)$ grows with $k$ increasing. While passing from $k = \overline{k} - 1$ to $k = \overline{k} + 1$ the function $C(\beta_k)$ has the sharp jump. Summing the information about $C(\beta_k)$ behavior, one can obtain the plot of $C = C(\beta_k)$ (Fig 1, solid line). In the discrete case, the obtained curve $C(\beta_k)$ is an approximation of that in the continuous case. The greater $n$ and $N$ values one takes, the more precise approximation is obtained.

Let us illustrate the proposed approach application to the Weierstrass function [15]:

$$W(t) = \sum_{n=-\infty}^{n=\infty} \frac{1 - \cos(b^n t)}{2^{(2-D)n}}$$

at $D = 1.5$. The behavior of $C(\beta_k)$ for the Weierstrass function is shown in Fig 2 (left panel). The values of $C(\beta_k)$ have been calculated for the first 30 points of the time series $X_i = W(i/N)$, $N = 1000$, $i = 1,...,N$ at $n = 100$. In Fig 2 one can see that $C(\beta_k)$ has a jump near the point $\beta_k = 0.5$. It is not surprising since it is well known [15] that the Holder exponent for the Weierstrass function is equal to $\alpha \equiv 2 - D$.

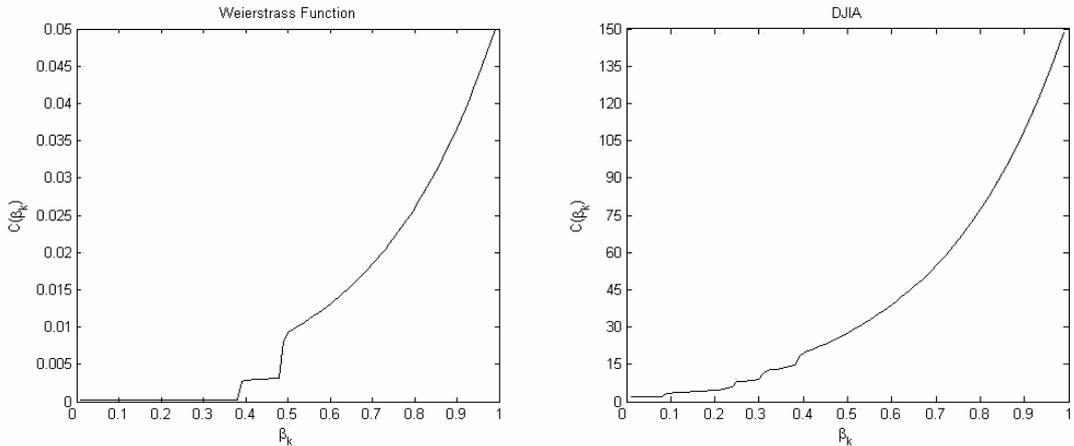

Fig. 2. The behavior of $C(\beta_k)$ for the Weierstrass function (left panel) and DJIA index (right panel).

In Fig 2 (right panel), the behavior of $C(\beta_k)$ for some financial time series is shown. This time series is composed from 30 daily close prices of Dow Jones Industrial Average (DJIA) index. One can see that it is practically impossible to determine exactly the point where the jump of $C(\beta_k)$ takes place. Hence, it is difficult to determine exactly the Holder exponent for financial time series in question. It means that the method not always is correct for financial time series for which their regularity is unknown a priory. To improve the approach a novel indicator based on the Holder semi-norm is introduced.

Let us consider the first $N'$ values $\{X_i\}_{i=1}^{N'}$ of the time series $\{X_i\}_{i=1}^{N}$. Suppose that for the time series $\{X_i\}_{i=1}^{N'}$ the Holder exponent is equal to $\alpha$. Knowing $\alpha$, let us calculate $C(\alpha)$ for the set $\{X_i\}_{i=1}^{N'}$ by means of the equation (3). Further for the set $\{X_i\}_{i=N'+1}^{N}$ the quantities $C(\beta_k)$ are calculated. But the procedure of calculations will be changed. In step by step calculating the numbers $C(\beta_k)$ let us stop at that moment when $C(\beta_k)$ runs up to the value of $C(\alpha)$. Let it happens at the time $\overline{k}$. Then one defines the quantity $\overline{a} = \beta_{\overline{k}}$. Thus for arbitrary $\alpha$ one can calculate $\overline{\alpha} = \overline{\alpha}(\alpha)$. The indicator $\overline{a}$ will be called the modified Holder exponent (MHE) of the set $\{X_i\}_{i=N'+1}^{N}$ with respect to the set $\{X_i\}_{i=1}^{N'}$.

Let us show that MHE is related to the Holder exponent by the example of the generalized Weierstrass function [22]:

$$V(t) = \sum_{k=0}^{\infty} 3^{-ks(t)} \sin(3^k t),$$

where $s(t)$ is some function with the range $(0,1)$. For this purpose let us remind [21] the notion of the Holder exponent at some point $t_0$.

*Definition 2. Function $f(t)$ has the Holder exponent $\alpha$ at the point $t_0$, iff*
1. *for any real $\gamma < \alpha$*

$$\lim_{h \to 0} \frac{|f(t_0 + h) - P(h)|}{|h|^\gamma} = 0$$

*and*
2. *if $\alpha < \infty$, for any real $\gamma > \alpha$*

$$\limsup_{h \to 0} \frac{|f(t_0 + h) - P(h)|}{|h|^\gamma} = +\infty,$$

*where $P$ is a polynomial of order not greater then $\alpha$.*

It is well known [22], that for the generalized Weierstrass function the Holder exponent $\alpha$ at the point $t$ has the form $a(t) = s(t)$ for any $t$. In Fig. 3 the generalized Weierstrass function when $s(t) = |\sin(5\pi t)|$, $t \in [1,2]$ is given (left panel).

MHE at the point $t$ for the generalized Weierstrass function on the interval $t \in [1,2]$ has been calculated as follows. This interval was divided into 100 subintervals with boundaries at points $t_k = k/M$, k = 101, ..., 200, $M = 100$. In calculating MHE at the point $t_k$ with a fixed number $k$ the set $\{X_i\}_{i=N'+1}^{N}$ was constituted from values of the generalized Weierstrass function taken at 7 points from the vicinity of the point $t_k$:

$$\{X_i\}_{i=N'+1}^{N} = \{V(t_{k-3}),\ V(t_{k-2}),\ ...,V(t_{k+3})\},$$

where $N = k+3,\ N' = k-4$.

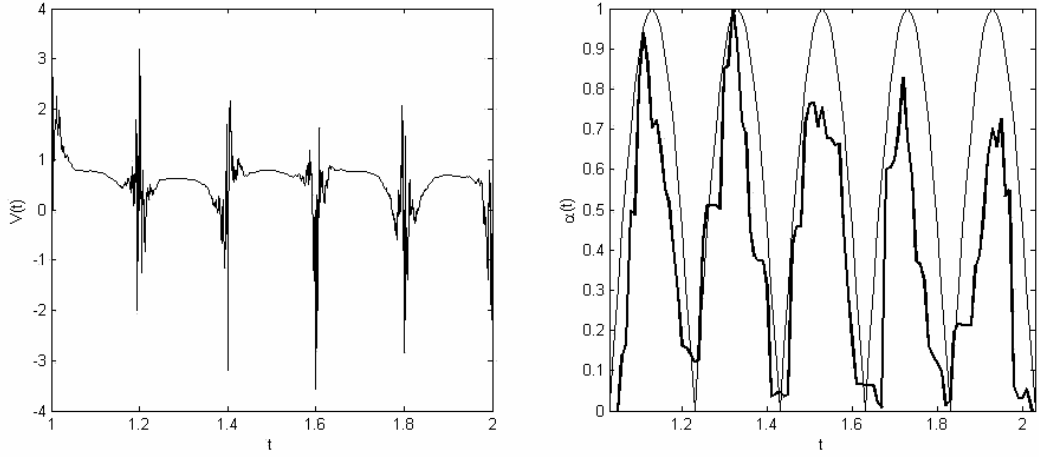

Fig. 3. The generalized Weierstrass function at $s(t) = |\sin(5\pi t)|$, $t \in [1,2]$ (left panel). The comparison of MPHE (heavy line) and theoretical values of the pointwise Holder exponents (thin line, right panel).

As the set $\{X_i\}_{i=1}^{N'}$ the values of the generalized Weierstrass function calculated at 35 points foregoing the point $t_{k-3}$ have been taken. The value of $\alpha$ has been fixed and equal to 1/2. Than one has to calculate the quantity e $C(1/2)$ for the set $\{X_i\}_{i=1}^{N'}$. However for reduction of computational resources and for smoothing of the MHE behavior the average $\frac{1}{5}\sum_{j=1}^{5} C_j(1/2)$ was calculated. Here $C_j(1/2)$ were calculated for values of generalized Weierstrass function at points of the consecutively arranged sets $E_j = [t_{k-3-7j},...,t_{k-3-7j+6}],\ j = 1,...,5$.

Let us call the MHE calculated at some point as the pointwise MHE (MPHE). In Fig 3 (right panel, heavy line) MPHE for the points $t_k$, $k = 101,...200$ which belong to the interval $t \in [1,2]$ are shown. In the same Fig 3 by the fine line the Holder exponents for the generalized Weierstrass function at $t \in [1,2]$ are also shown. The values of MPHE have been normalized since at the points of local maximum their values were greater then 1. However, the character of increasing and decreasing of the Holder exponents and of MPHE looks very similar.

**4. TESTING ON FINANCIAL DATA**

For financial time series MPHE are calculated almost in the same manner as for the generalized Weierstrass function. The choice of the set $\{X_i\}_{i=N'+1}^{N}$ is the only difference. The values $\{X_i\}_{i=N'+1}^{N}$ are taken at points located only to the left from the point $t_k$. Namely, $\{X_i\}_{i=N'+1}^{N} = \{X(t_{k-w+1})...,X(t_{k-1}),X(t_k)\}$, where $N = k$, $N' = k-w$ and $w$ is a parameter described below. Due to the fact that MPHE computation does not involve the points located to the right of the point $t_k$ and treated as a future which is unknown the indicator MPHE can be used as a predictor. Moreover this indicator does not have boundary effects, i.e. while adding new values of the considered time series the heretofore calculated values of MPHE indicator remain the same. The number $w$ of time points defining the set $\{X_i\}_{i=N'+1}^{N}$ is a parameter in the MPHE calculations. This parameter is called in further "the window size".

**4.1 Prediction of critical points in stocks quotations**

Here the critical points are treated as trends change, crashes and sharp quotations increasing. The prediction is realized by allocation of the time intervals where MPHE demonstrate sharp increasing or in other words the bursts. In order to single out the significant bursts the so-called "signal line" is used. The signal line $sl(t)$ is calculated as the sum of average of MPHE in previous to $t$ time period and $k$ standard deviations in the same time period. Namely:

$$sl(t) = \overline{G}(t) + k \left( \frac{1}{h-1} \sum_{i=0}^{h-1} \left( \overline{G}(t) - G(t-i) \right)^2 \right)^{1/2}, \overline{G}(t) = \frac{1}{h} \sum_{i=0}^{h-1} G(t-i),$$

where $G(t)$ is the value of MPHE at the time $t$. The numbers $k$ and $h$ are parameters of the proposed approach. The prediction is defined as the occurrence of an event when MPHE cross the signal line bottom-up. In such cases it is assumed that there is a signal for the prospective critical point in financial time series. The parameter $k$ will be called "the signal line height" and the value of this parameter strongly influences on signals appearance. The greater is the value of $k$ the less is the number of signals. At the same time when the value of $k$ decreases a number of the so-called "false" signals also decreases. By definition the signal is false if after signal appearance there are no critical points in future. The parameter $h$ which is called "the history of signal line" is usually proportional to the window size with the coefficient of proportionality equal to 10. This value of the coefficient of proportionality has been optimized in numerical experiments.

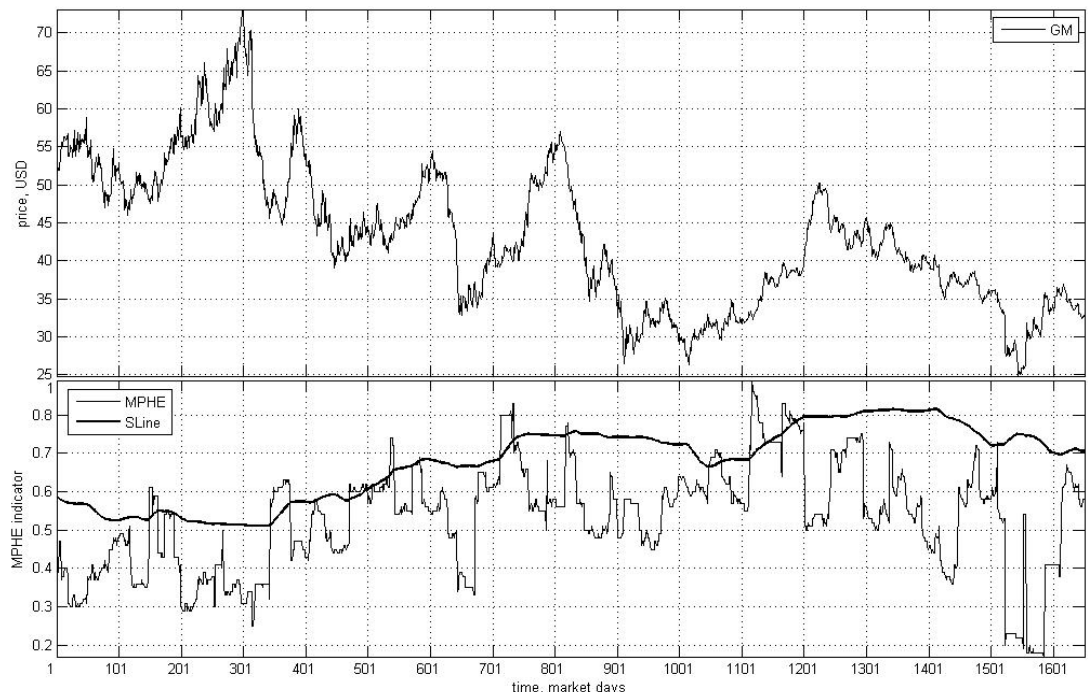

Fig. 4.  Prediction of the critical points for General Motors stocks quotations. In the upper plot the daily close prices in the period 1999 - 2005 are presented. In the power plot the MPHE indicator and the signal line are shown.

For the testing of the proposed approach the time series consisted from the daily close prices of the General Motors stocks quotations in the period from February 25, 1999 till September15, 2005 (1650 daily observations) has been chosen. The data have been taken from http://finance.yahoo.com/ . For the chosen time series the MPHE have been calculated at the parameters values $w=30$, $k=1.5$ (see Fig. 4) which have been determined using the small part of 1650 daily observations. As the result the most of the MPHE local maxima coincide with the local extremes of the time series or precede the time series large movements (not less then 20%). In that way after the most MPHE signals one of the predicted events occurs: trend change, large take-off or decline. The prediction for different events takes place in average 1-2 months prior to the event beginning, i.e. the MPHE indicator crosses the signal line in average 20-40 days prior to the critical event beginning. In frames of the proposed approach more exact estimations are impossible since it is very difficult to define in a formal way when the beginning of a critical event.

**4.2 Crashes of DJIA index prediction**

During the last 70 years for the DJIA index there have been picked out 9 critical events (see Figs. 5-6). Among these critical events there are well known historical crashes such as October 1987, August 1998, September 2001, as well as other critical events (November 1937, March 1938, May 1940, May 1962, May 1974, August 2002). All of these events are characterized by strong decline in short period of time (not less then 9% in three days). For these events analysis the time series consisted from the daily close prices of DJIA index was considered, the MPHE indicator and the signal line were calculated. The model parameters have been taken the same as for the General Motors stocks quotations analysis with the exception of the signal line height. The height was reduced to 1. The reduction of this parameter is determined by the fact that volatility of the DJIA index is less than volatilities of their components. Almost for all cases there is the "reliable" MPHE signal before the critical event that means that almost all events have been predicted. The prediction horizons changed within the bounds of 100 days. The "reliable" MPHE signal is treated as follows. Either MPHE signal occurs directly before the crash and continues till the crash or the MPHE signal continues till the point of trend change which foregoes a crash. There are also the MPHE signals which occur in due time before the crash and continue right up till the crash. Only in the case of the August 1998 crash there is no "reliable" MPHE signal, but the situation is very similar to the successfully predicted the May 1940 crash.

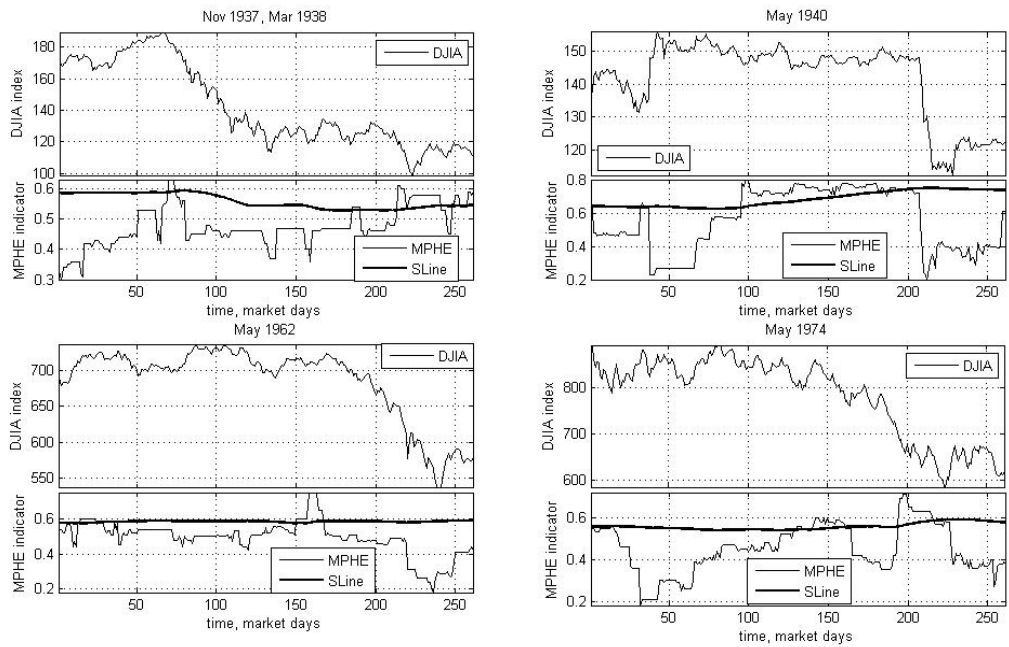

Fig. 5.    Prediction of the DJIA index in the period from 1937 till 1974 years.

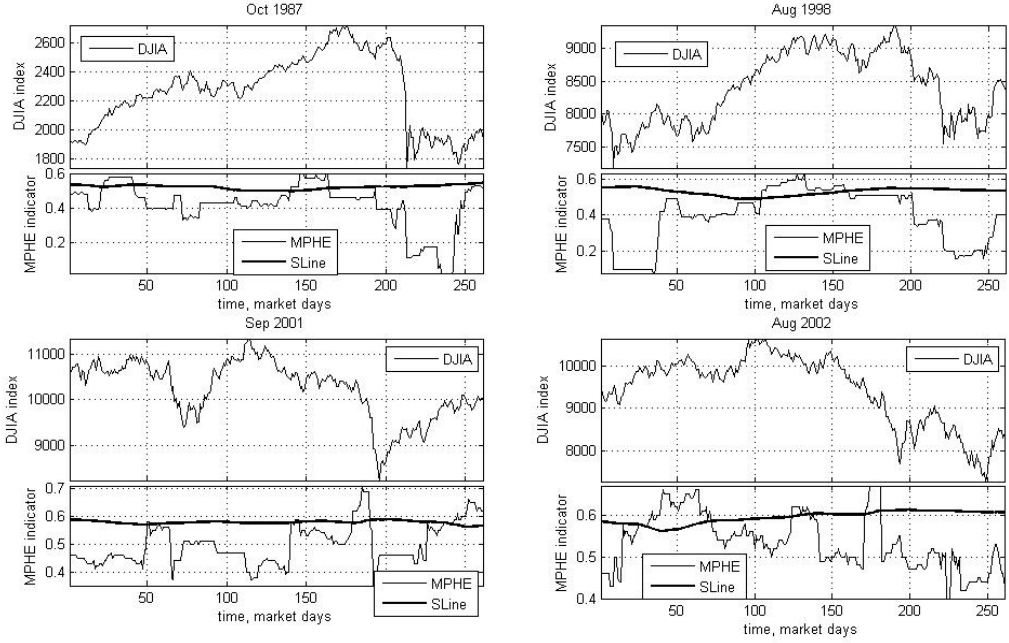

Fig. 6.    Prediction of the DJIA index in the period from 1987 till 2002 years.

**4.3 Stock market crashes prediction**

In order to improve the prediction quality of the market critical points in whole it is proposed to elaborate an indicator not for the market index but for the financial instruments from which it is composed. Let us introduce the so-called "joint modified pointwise Holder exponent" (JMPHE) for all 30 stocks which constitute the DJIA index. More generally let us consider the set of time series $X_i(t), i = \overline{1..n}$, the calculated MPHE $G_i(t)$ and the corresponding signal lines $sl_i(t)$. Then the JMPHE $H$ for the set of time series $\{X_i(t)\}_{i=1}^{n}$ is defined as follows:

$$H(t) = EMA\left(\frac{1}{n}\sum_{i=1}^{n} sign(G_i(t) - sl_i(t))\right).$$

Here $EMA$ is the exponential moving average. Let us remind that for arbitrary time series $x(t)$, $t = 1, 2, ...$ the $EMA$ indicator with the parameter $\lambda$ is given by

$$EMA(t) = x(t)\lambda + EMA(t-1)(1-\lambda).$$

In calculations of the present paper the $EMA$ indicator is used with the parameter $\lambda = 0.3$. Thus the JMPHE indicator has the meaning of the counter of stocks which produce a signal at the given moment of time. In the preset paper the signal line for JMPHE has been chosen as the constant equal to $1/2k$.

The proposed approach within the period 1999-2005 gives from 1 up to 10 signals (predictions) at the various values of the parameter $k$ and the fixed value of $w = 30$. In Fig 7 the plot DJIA index within the period 1999-2005 (1650 daily close prices, upper panel), JMPHE indicator at $k = 1.5$ (middle panel) and JMPHE indicator at $k = 2.5$ (lower panel) are shown. By the control of the parameter $k$ it is possible to cut off some signals. Thus at $k = 1.5$ there are 7 signals, whereas at $k = 2.5$ there remains only one signal which predicts the September 2001 crash prior 2 months before this crash. It should be noted that signal shapes before the trend changes and crashes are different. Before crashes the shape of JMPHE indicator looks like sharp bursts (see, for example, signals before September 2001 crash near the point # 600 in Fig. 7, middle panel). Before the trend changes JMPHE indicator increases slowly (see, for example, the signal at the 1999 year near the point # 100 in Fig. 7, middle panel).

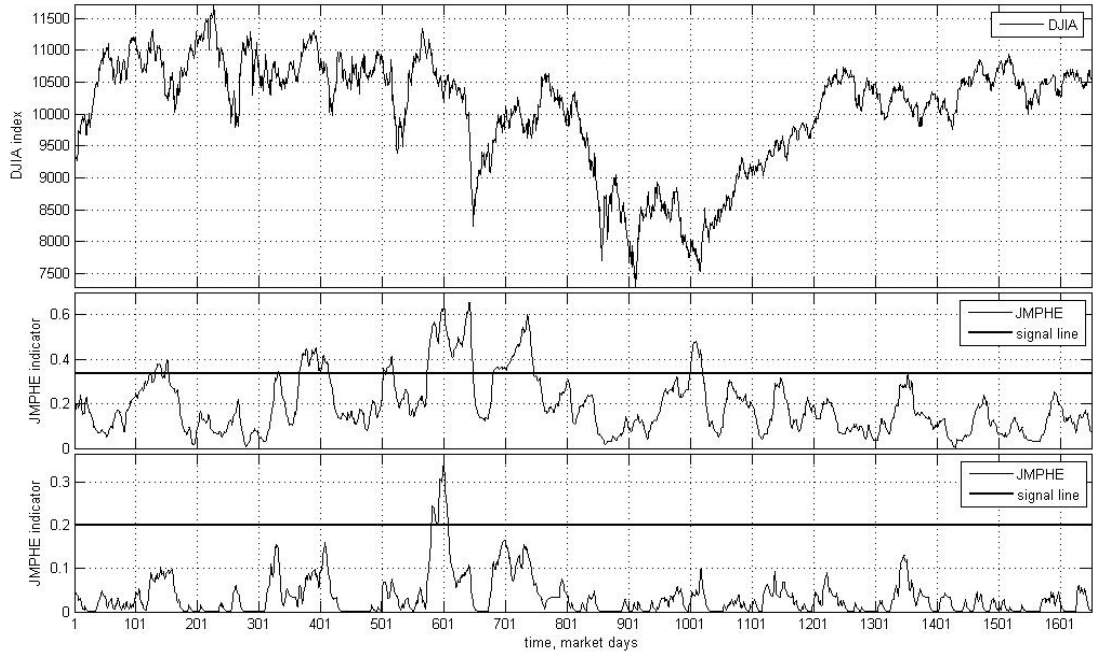

Fig. 7.   Prediction of the USA stock market critical points in the period 1999 - 2005.

**4.4 A trading strategy based on JMPHE and VIX**

Using the JMPHE indicator for the market behavior predictions and volatility index VIX (http://www.cboe.com/micro/vix/VIXoptionsQRG.pdf ) a novel trading strategy has been elaborated. In frames of the strategy the JMPHE indicator is used for large movements prediction and hence for the choice of the proper time for the investment decision making. Since the JMPHE indicator does not determine the direction of the market movement the index VIX was used for the solution of this problem. Elaborated by CBOE the VIX index is calculated on the basis of "out-the-money" options on the S&P 500 index and gives the possibility to determine the direction of future market movement (see Fig. 8). In the upper panel the plot of S&P 500 index(1650 values of daily close prices within the period 1999-2005) is shown. In the lower panel the plot of the VIX index (1650 values at the market close within the period 1999-2005) is also shown. The VIX data have been taken from www.cboe.com. One can see that the local VIX maxima coincide with the local S&P500 minima and vice versa. It means that one can use the VIX index for identification of the market movement direction. If VIX is near the local maximum then market is close to the local minimum and hence will increase. The inverse statement is also true. For the determination of the minimal and maximum values of VIX two boundary values 20 and 30 have been introduced. If VIX<20, then it is assumed that VIX has the minimum, i.e. there is a signal to open long position. If VIX crosses the value 30 bottom-up there is a signal to open short position.

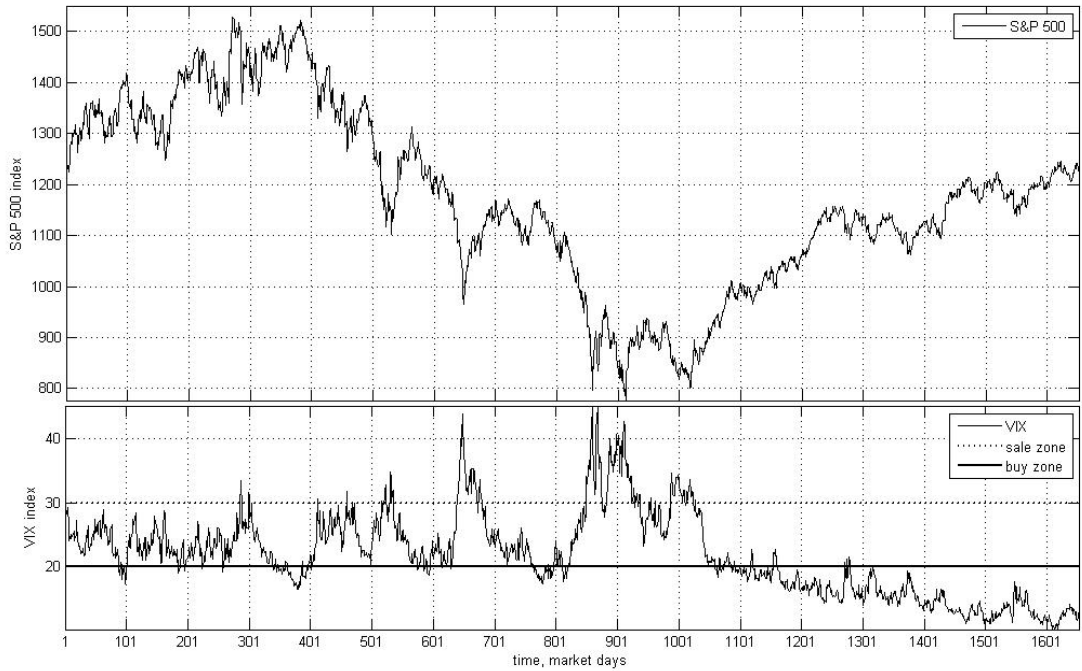

Fig. 8. VIX index for S&P 500 in the period 1999 - 2005.

In the proposed strategy the portfolio S&P500 is traded. At the same time the JMPHE indicator is calculated for the stocks from the DJIA portfolio at the parameters values $w = 30$ and $k = 1.5$. It is justified by the fact that DJIA and S&P500 indexes are strongly correlated and hence using DJIA instead of S&P500 one can essentially reduce the time of calculations. The rules for open/close position are the following. If there is the JMPHE signal and during this signal or during 30 days after the JMPHE signal there appears also the VIX signal then position is opened. The VIX signal delay in 30 days with respect to JMPHE signal is acceptable due to the following reasons. Firstly, the JMPHE signals begin always before the VIX signals. Secondly, the time period in 30 days has been chosen since this value coincides with the average horizon prediction (see Section 4.3). If the next JMPHE signal begins the previously opened position is to be closed. If the short position has been opened then it is closed when the VIX indicator gives the signal to open long position. On the contrary if the long position has been opened it remains opened even if the VIX indicator gives the signal to open short position. It is explained by the fact that VIX location near its lower boundary is not reliable indication that the market is in the local minimum as it was, for example, during 2004 and 2005 years (see observations 1200-1650 in Fig. 8).

The proposed trading strategy has been tested in the period from February 25, 1999 till September15, 2005. The first trading day when all strategy conditions for opening position were satisfied occurs on July 7, 2000. Altogether the market positions were opened and closed 6 times during the testing period. The triaging was realized with the reinvestment of the strategy profits (see Fig. 9 where the strategy capital behavior and dynamics of S&P500 index are shown). The commission was not taken into account. September 15, 2005 was the last trading day. In total during 5 years and 2 months the gross profit of the strategy constituted 156%. The maximal capital drawdown was equal to 8%. It is much better than -15% as market profitability during testing period and 49% as maximal drawdown. Here the market profitability is treated as the profitability of the "buy&hold" strategy.

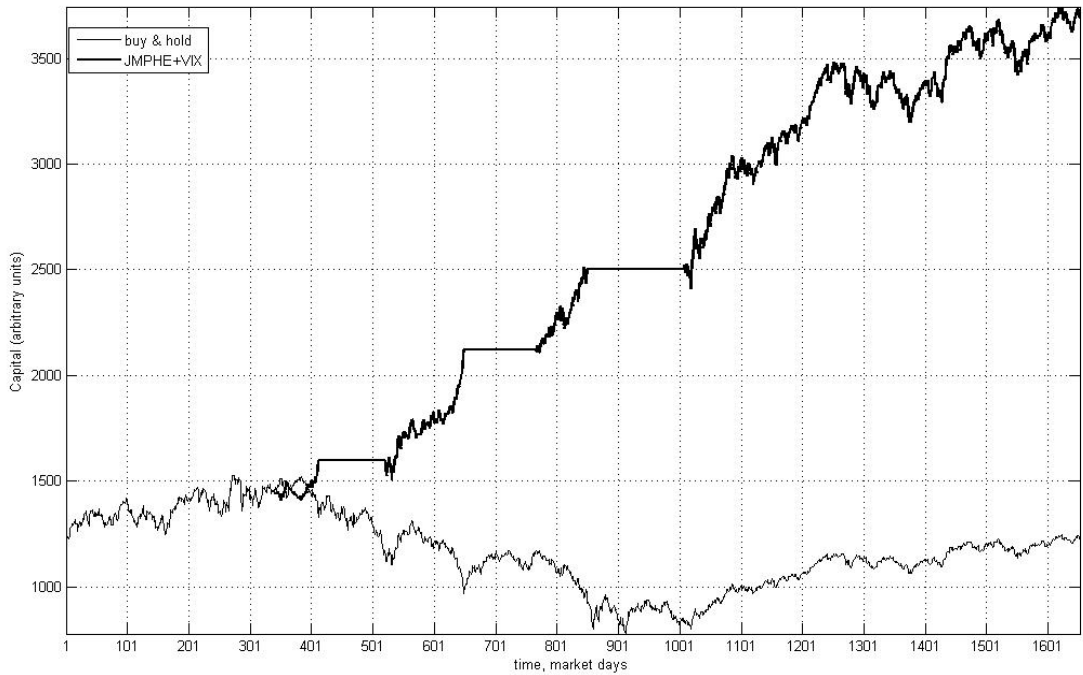

Fig. 9. Capital of the strategy based on the combination of MPHE indicator and VIX index (heavy line). The behavior of S&P500 index (thin line).

## 5. SUMMARY

A novel approach to the modified Holder exponents definition and numerical algorithm for their calculation have been elaborated. By means of the proposed approach it is possible to calculate the pointwise Holder exponents for discrete time series. This method has been tested on the stylized data. It has been shown that the calculated and theoretical results are in fair agreement.

The well known hypothesis [12-14] concerning the regularity increasing of financial time series before market critical events has been numerically verified. The proposed approach differs from one elaborated by the Groupe Fractales and realized in Fraclab package. In contrast to the local Holder exponents calculations which are realized in Fraclab our calculations used only the points of time series lying to the left of the present time point. It means that the elaborated MPHE and JMPHE indicators can be used as predictors since they do not use the information from the future.

It has been shown that the proposed approach is efficient in the real world applications, namely for the 30 financial time series from DJIA index. By means of MPHE and JMPHE indicators the critical points in General Motors stocks quotations in the period 1999-2005 years as well as most crashes in DJIA index during last 70 years have been predicted. On the basis of joint modified pointwise Holder exponents constructed for the "blue chips" of the USA stock market the flexible forecasting system for prediction of the critical events and crashes in the USA stock market has been elaborated. The testing of this system in the period 1999-2005 showed that for different values of the parameters it gives from 1 to 10 signals before the most significant critical events in this market. When the parameters values are varied there remained only one signal before the September 2001 crash which was the most significant event in the tested period. It should be noted that the signal appearance before the September 11, 2001 crash indicates that the market already contained precursors of crash and the act of terrorism at September 11, 2001 was just a trigger for the crash followed.

On the basis of the market critical points predictions and their combination with the VIX indicator there has been elaborated the trading strategy with profitability more than 150% in 5 years at 8% as the maximal drawdown. The proposed trading strategy is much more profitable than the "buy&hold" strategy.

## 6. ACKNOWLEDGMENTS

Authors are indebted to D. Latypov and L. Dmitrieva for many fruitful discussions and to RD Partners Texas, LLC for encouragement.